\documentclass{aastex}
\usepackage{epsfig,amsmath,amsfonts,amssymb,graphicx,subfigure,enumitem}

\usepackage{tabu}
\newcommand{\uvec}[1]{\boldsymbol{\hat{\textbf{#1}}}}
\renewcommand{\arraystretch}{1.0}
\begin{document}
\title{Propagation of MRT Unstable Plasma Spikes in the Inter-planetary Space}
\author{Sudheer K.~Mishra}
\affil{Department of Physics, Indian Institute of Technology (BHU), Varanasi-221005, India.}

\author{Talwinder Singh}
\affil{Department of Space Science, University of Alabama in Huntsville, Huntsville AL 35899,USA}

\author{P.~Kayshap}
\affil{UMCS, Group of Astrophysics, UMCS, ul. Radziszewskiego 10, 20-031, Lublin, Poland}

\author{A.K. Srivastava}
\affil{Department of Physics, Indian Institute of Technology (BHU), Varanasi-221005, India}

\begin{abstract}
We have used the Coronagraphic and Heliospheric Imaging data from Solar TErrestrial RElations Observatory (STEREO) to observe a prominence which is erupted on June 7$^{th}$ 2011. This prominence is subjected to the morphological evolution of MRT instability from the lower solar corona upto the inter-planetary space. The unstable structures are converted into the bunch of localized plasma spikes due to the turbulent mixing, and propagate in the inter-planetary space upto 1 A.U.
\end{abstract}

\vspace{-1.0cm}
\section{Introduction}\label{sec:intro}
The formation of plumes and fragmentation in an eruptive prominence may be occurred due to the magnetic Rayleigh-Taylor instability (Hillier 2018). A prominence eruption occurred on June 7$^{th}$ 2011 as observed by Solar TErrestrial RElations Observatory (STEREO). It is observed that the eruptive prominence is magnetic Rayleigh-Taylor unstable from the inner corona to the low interplanetary space (Innes et al. 2012; Carlyle et al. 2014; Mishra et al. 2018). We demonstrate that the eruptive prominence eruption moves in form of the MRT unstable plasma spikes into the inter-planetary space reaching upto 1 A.U. (Wood et al. 2015).
\section{Observational Data and Analysis}
Coronagarphs (COR-1A \& COR-2A, Howard et al. 2008) and Heliospheric Imagers (HI-1A \& HI-2A, Howard et al. 2008) onboard STEREO have been used to observe the prominence eruption occurred from the NOAA AR 11226/11227 on June 7$^{th}$ 2011 from 1.4 solar radii to 1 A.U. The MRT unstable
plasma structures have been observed from inner corona upto the low interplanetary space (Innes et al. 2012; Mishra et al. 2018). Here, we extend the study of Mishra et al. (2018) and report that MRT unstable plasma spikes associated with prominence eruption propagate into an outer inter-planetary space.
\begin{figure*}
\begin{center}
\caption{Right: Partial field of view of COR-1 onboard STEREO-A that shows the MRT unstable fingers in the eruptive prominence at June 7$^{th}$ 2011. Left: A schematic representation of a MRT unstable finger structure. 
}
\includegraphics[width=5.0in]{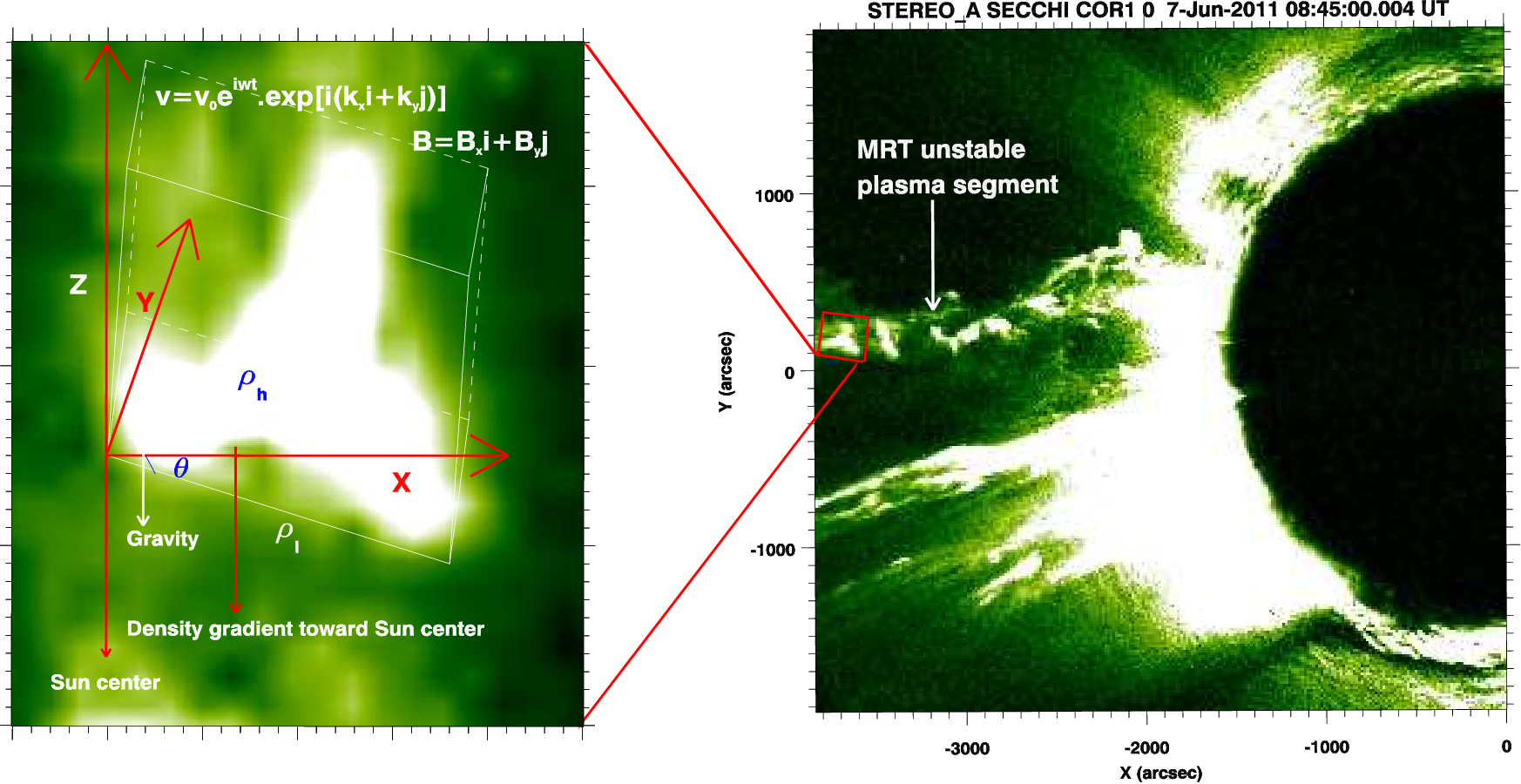} 
\end{center}
\end{figure*}
\begin{figure}
\begin{center}
\caption{Left: 
The aligned and composite images of STEREO/COR-2A (2.5 to 15 solar radii), Heliospheric Imager (HI)-1A (15 to 80 solar radii), and Heliospheric Imager (HI)-2A (upto 4 A.U.) onboard Solar TErrestrial Relations Observatory (STEREO) on 9 June 2011 at 12:09 UT, display the formation of plasma spikes impinging in and around the Earth.
Right: A schematic showing the propagation of MRT unstable plasma spikes in the inter-planetary space.}
\mbox{
 \includegraphics[width=3.5in]{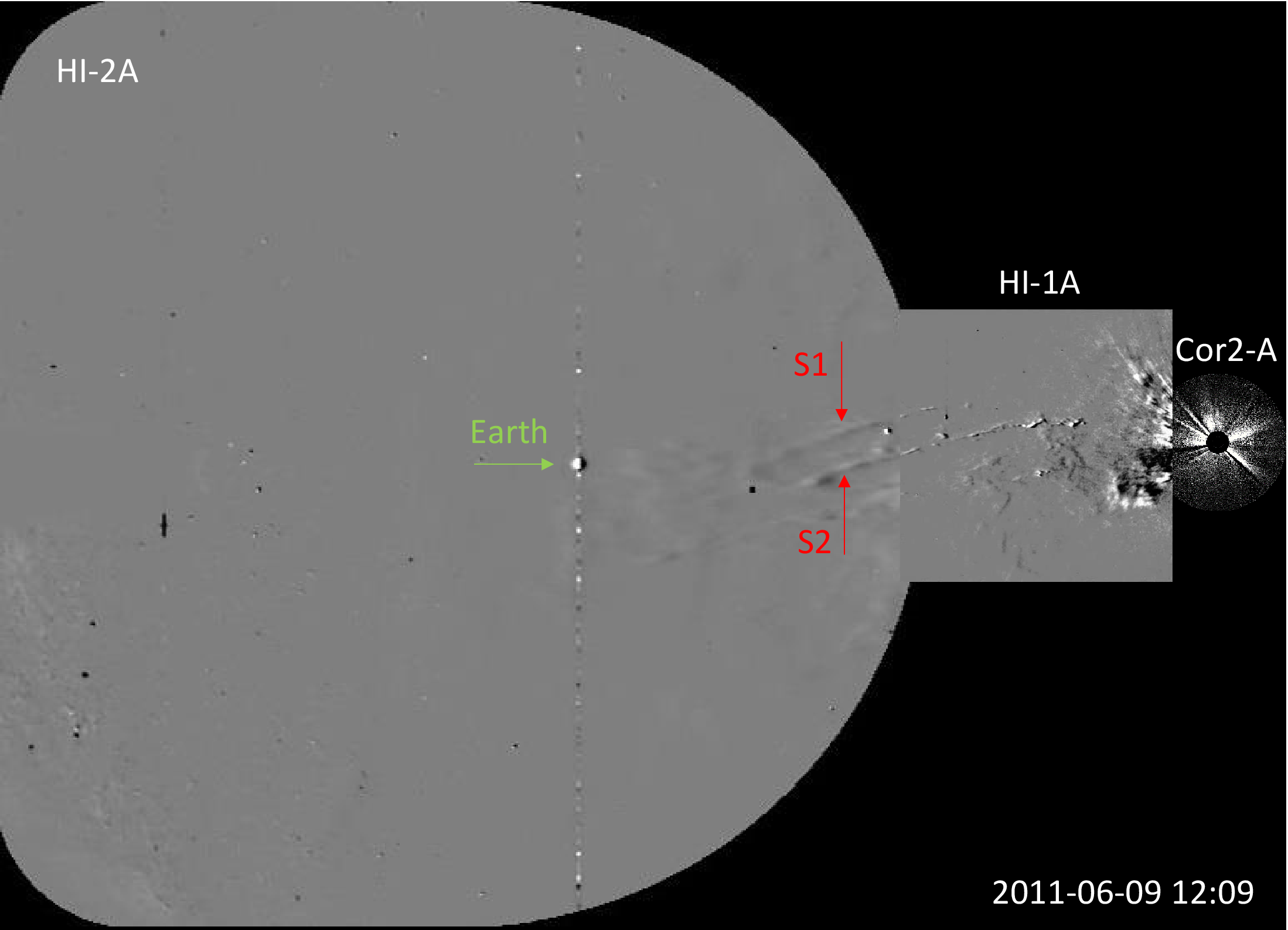}
\includegraphics[width=2.55in]{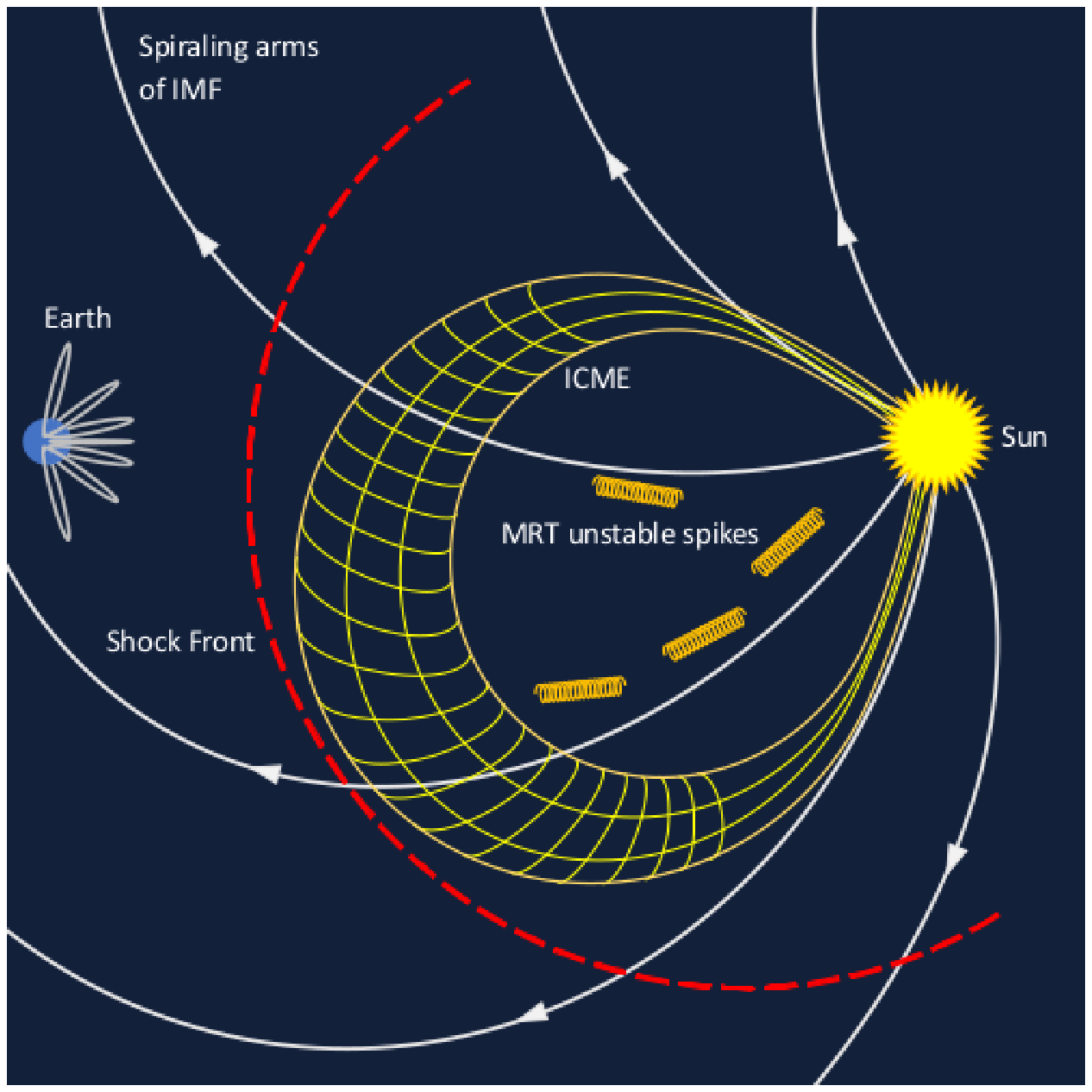}}
\end{center}
\end{figure}
\section{Result and Discussion}
A prominence has been erupted on June 7$^{th}$ 2011 and some of the associated plasma have been reached upto 1 A.U. on June $9^{th}$ 2011 (Wood et al. 2015). The magnetic Rayleigh-Taylor unstable finger structures have been observed into the intermediate corona using STEREO/COR-1 (Fig.~1; right panel). Schematic in Fig.~1 (left-panel) shows a planer magnetic interface $(B=B_{x}\uvec{i}+B_{y}\uvec{j})$ developed between the finger structure and corresponding below lying dark region. The magnetic interface is parallel to the wave velocity perturbation triggering MRT instability. A density gradient works towards the Sun center in the direction of the gravity (Fig.~1; left panel). The tension component of the Lorentz force accelerates the MRT unstable plasma segment against the gravity and density gradient (Fig.~1; left panel). It is observed that these MRT unstable finger structures further change there morphological shape i.e., fingers$\rightarrow$ mushroom-like $\rightarrow$ plasma spikes, with the decreasing magnetic field into the outer solar atmosphere. (Mishra et al. 2018). \\
\vspace{-0.1cm}
The plasma spikes are evolved in the low-interplanetary space due to turbulent mixing of the MRT unstable plasma structures erupting from the Sun (Mishra et al. 2018). The transport of this eruptive prominence upto 1.0 A.U. has reported by Wood et al.(2015). In the present paper, we find that the eruptive prominence is MRT unstable which propagates in form of plasma spikes into the inter-planetary space (cf., Figs~2, left-panel, spikes S1\& S2). The schematic (Fig.~2, right-panel) demonstrates the modified scenario of the propagation of various ejecta in the interplanetary space. It includes the formation of localized MRT unstable plasma spikes that can slide along the inter-planetary magnetic field and can carry their own magneto-plasma system. These unstable plasma spikes imping around the Earth at 1 A.U. (Fig.~3).
If such MRT unstable plasma structures move collectively and interact with the Earth's outer atmosphere then they can generate the episodic geomagnetic storm and can act as a new space weather candidate.

\end{document}